\documentclass[twoside,slac_one]{revtex4}
\usepackage{graphicx}
\usepackage{fancyhdr}
\usepackage{amsmath} 
\usepackage{bm}
\usepackage{amsxtra}
\usepackage{amssymb}
\usepackage{amsthm}
\usepackage{latexsym}
\usepackage{lscape}

\newcommand{\ket}[1]{\vert #1\rangle}
\newcommand{\bra}[1]{\langle #1\vert}

\begin{document}

\title{A novel algorithm for computing quark propagators for lattice hadron spectroscopy}

%

\author{J. Bulava}
\affiliation{NIC, DESY, Platanenallee 6, D-15738, Zeuthen Germany}
\author{J. Foley}
\affiliation{Department of Physics, University of Utah, Salt Lake City, UT 84112, USA}
\author{Y.C. Jhang}
\affiliation{Department of Physics, Carnegie Mellon University, Pittsburgh, PA 15213, USA}
\author{K.J. Juge}
\affiliation{Department of Physics, University of the Pacific, Stockton, CA 95211, USA}
\author{D. Lenkner}
\affiliation{Department of Physics, Carnegie Mellon University, Pittsburgh, PA 15213, USA}
\author{M.J. Peardon}
\affiliation{School of Mathematics, Trinity College, Dublin 2, Ireland}
\author{C.J. Morningstar}
\affiliation{Department of Physics, Carnegie Mellon University, Pittsburgh, PA 15213, USA}
\author{C.H. Wong}
\affiliation{Department of Physics, Carnegie Mellon University, Pittsburgh, PA 15213, USA}

\begin{abstract}
A novel algorithm for estimating quark propagation for lattice hadron spectroscopy calculations will be presented. Lattice QCD is a framework in which the nonperturbative, first principles computation of hadronic correlation functions is possible by numerical simulations. It requires a lattice regulation of QCD in a finite Euclidean space-time where correlation functions are evaluated numerically via Monte-Carlo importance sampling methods. The asymptotic behaviour of correlation functions in Euclidan time is then used to extract the energies of the hadronic states of interest. 

This approach of computing correlation functions becomes numerically very challenging as one attempts to evaluate multi-hadron correlators and disconnected diagrams due to the rapid increase in the number of quark propagators involved in the calculation. Each quark propagator is obtained from an inversion of the 4-dimensional Dirac matrix in a finite, but large volume. Disconnected diagrams (quark loop diagrams) are particularly problematic because they require quark propagators from every point on a timeslice to every other point on the same timeslice, for all timeslices of the lattice.

The cost of inverting the Dirac matrix, in terms of CPU cycles, increases as the quark mass being simulated approaches the physical values and the space-time volume is enlarged to reduce the finite size effects. It is not practical to compute these ``all-to-all" quark propagators with the current resources available and this  limits the physics that can be addressed in Lattice QCD.

The standard solution is to stochastically estimate the all-to-all propagators with random noise sources. This method, however, introduces a lot of noise into the calculation which is reduced by performing more and more inversions. The stochastic LapH algorithm for inverting quark propagators avoids "using noise to cancel noise" approach by first reducing the space by cutting out the high frequencies (LapH) and then introducing noise sources in the LapH subspace. 

The usual volume scaling problems of all-to-all algorithms are absent in this method and the judicious choice of noise-dilution schemes makes the algorithm practical for real simulations that are performed near physical values of the quark masses. The signal for selected correlation functions will be shown to demonstrate the efficacy of the algorithm.
\end{abstract}

\maketitle

\section{Introduction} There has been significant progress in lattice QCD simulations in recent years now that dynamical simulations can even be done at physical quark masses (\cite{Durr:2008zz}, \cite{Aoki:2009ix}). Calculations of level splittings due to electromagnetic corrections (see review \cite{Colangelo:2010et} and references therein) have also been attempted in recent years, as well as realistic attempts to calculate binding energies of atoms (\cite{Inoue:2010es}). The computation of the excited state spectra of baryons and mesons has been ongoing for the single particle sector in both $N_f=2$ and $N_f=2+1$ dynamical calculations at relatively large quark masses. The Hadron Spectrum Collaboration has been developing the tools necessary to complete this analysis starting with the design of the lattice interpolating operators for single particle states (\cite{Basak:2007kj}).

The question of mixings of excited single-particle states with multi-particle states has been pointed out in selected channels in recent spectroscopy papers (see for example \cite{Bulava:2009jb}, \cite{Aoki:2009ix}, \cite{Dudek:2011tt} and references therein), but these studies do not see multi-particle states in their spectrum even where they are expected (given the parameters of the simulations). The main problem appears to be similar to the situation of the absence of string-breaking of the static potential in dynamical simulations without explicit two-meson operators in the variational basis. 

In these proceedings, we report on the progress made in filling in this void by using an estimate of all-to-all quark propagators via the {\em stochastic LapH} algorithm introduced earlier this year (\cite{Morningstar:2011ka}). We apply the algorithm to simple two-particle spectra with finite momenta hadron operators and to correlation functions which have disconnected diagram contributions, such as those in the isosinglet channel, including diagrams of mixing with scalar glueball operators.

\section{Spectroscopy in Lattice QCD}
Spectroscopy in lattice QCD is conventionally carried out by computing hadron correlation functions and fitting the exponential decay of the signal in Euclidean time:
\begin{eqnarray}
\bra{0}\mathcal{O}(t)\mathcal{O}(0)\ket{0}\rightarrow A_0e^{-E_0t}+A_1e^{-E_1t}+\cdots.
\end{eqnarray}
The interpolating operator $\mathcal{O}$ only needs to have a finite overlap with the state in question in the infinite time ($t\rightarrow\infty$) limit for the extraction of the ground state. An example of such an operator would be the single pion operator for which $\mathcal{O}=\bar\psi\gamma_5\psi$ is usually sufficient for extracting the pion mass. However, signal-to-noise ratio deteriorates rapidly for heavier states with increasing time so that interpolating operators with better overlaps are preferred to determine the energies with reasonable errors. 

The hadronic correlation functions are computed by contracting the appropriate combination of quark propagators to project out the required lattice irreps of the states being sought. It is the computation of these quark propagators (which requires the inversions of a large, four-dimensional matrix) that is the most time-consuming part of the analysis at light quark masses. It is usually not feasible to compute all elements of the quark propagator at the quark masses that we are interested in, even with the largest computational resources available to us today. Therefore, it is usually the case that only a handful of quark propagators are computed for spectroscopy calculations.

The excited states in a given symmetry channel can be obtained by enlarging the interpolating operator basis and using the variational method to determine the optimal combination of those operators that project out the higher lying states. A correlation matrix is required in this case and effective interpolating operators must be included in the basis for this procedure to produce results with reasonable errors. For a two-particle state, it is preferred or even necessary that explicit two-hadron operators with finite momenta are included the basis of operators.

Single-particle states with disconnected contributions in the correlation functions
have the same problem as multi-particle states as quark propagators from many points on the lattice have to be computed to reduce the statistical errors that are typically intrinsically larger than connected correlation functions. The main problem comes from so-called ``$t$-$to$-$t$" diagrams where a quark propagates out from timeslice $t$ and propagates back to $t$ again. The efficient evaluation of these diagrams is a necessary step towards the determination of the low-lying hadron spectrum from lattice QCD. The stochastic LapH algorithm was mainly designed to solve this problem. Details of the stochastic LapH algorithm are given in \cite{Morningstar:2011ka}. We outline the basic idea and its implementation for hadron spectroscopy calculations in the following sections.  

\section{The Stochastic LapH Algorithm}
The stochastic LapH algorithm is based on two main ideas that have been found to be effective in reducing noise in the single timeslice-to-all and all-to-all propagators. The first is the idea of propagating only the low-lying modes of the quarks, as they are the modes which dominate the hadron correlation functions. These quark propagators go by the name of {\em LapH} propagators (or distilled propagators, ~\cite{Peardon:2009gh}). LapH propagators are a particular implementation of the distillation algorithm with a {\em H}eaviside step function cutoff of the eigenmodes ($v^{(i)}(\vec{x},t)$) of the the {\em Lap}lacian operator. The main object which takes the place of the smeared quark propagators (called perambulators in ~\cite{Peardon:2009gh}) take the following form in this case:
$\bf{V}^\dagger(t)M^{-1}_{\alpha\beta}(t,t_0)\bf{V}(t_0)$ where $\bf{V}(t)$ is the matrix formed from the column of eigenvectors of the Laplacian on timeslice $t$ and $M^{-1}_{\alpha\beta}$ propagates the eigenmodes on timeslice $t_0$ to $t$ (and spin component $\beta$ to $\alpha$). We note that the perambulators enables one to inject finite momenta into the quarks but still connects timeslice $t_0$ to $t$ in the same way that the usual point-to-all propagators do, which can complicate the formation of hadronic correlation matrices. Moreover, the evaluation of $t$-$to$-$t$ diagrams would require the computation of the perambulators from many timeslices on the lattice.

The second ingredient in the algorithm is the stochastic estimation of the quark propagators from all of the timeslices to make this a true all-to-all algorithm without computing all of the perambulators on the lattice. The use of stochastic noise to reduce the computational cost of all-to-all propagators is nothing new as it has been used in the computation of condensates and other quantities that have disconnected diagrams. In the stochastic LapH algorithm, $Z_4$ noise is introduced in the LapH subspace in order to keep the noise out of the eigenvectors that were computed ``exactly". This means that each eigenvector of the Laplacian that we are using acquires a single noise factor instead of the usual $N_{space}\times N_{color}\times N_{spin}$ factors, greatly reducing the unwanted effects of the stochastic method. The noise vectors introduced in this way are denoted by $\varrho^{[\alpha k \tau]}(\vec{x},t)$, where $\alpha$ denotes the eigenvector index, $k$ denotes the spin index and $\tau$ denotes the time index. The solutions of the Dirac equation with these noise sources carry similar indices, $\varphi^{[\alpha k \tau]}(\vec{x},t)$. 

The final ingredient in this algorithm is the use of dilution on the noise vectors for each quark line instead of averaging over a large number of independent sources to cancel the noise (see review in ~\cite{Peardon:2002ye} and references therein). This approach is much more effective than using multiple noise sources if one can find the right dilution scheme since the hopping of noise from one source to another is effectively removed with a judicious choice of the dilution scheme. Procedurally, the projection operators $P^{(d)}$ are applied to the $Z_4$ noise sources which dilutes the noise according to some scheme before they are used as the quark source in the Dirac equation. The diluted noise sources are denoted by, $\varrho^{[d]}(\vec{x},t)=P^{(d)}\varrho(\vec{x},t)$, where $d$ is the dilution index and a similar index is used for the corresponding solution.

The estimate of the full quark propagator for a given quark-line can then be written in the following compact form:
$$Q^{r}_{uv}=\sum_d\varphi^{[d]}_u(\rho^r)\otimes\varrho^{[d]\dagger}_v(\rho^r)$$
where the indices $u$ and $v$ are collective indices to simplify the notation and $r$ denotes the particular $Z_4$ noise source. A pion propagator, for example, can be formed from these propagators by performing the following contractions: $$C_\pi(t,t_0)={\rm Tr}\left[Q^{1}\gamma_5Q^{2\dagger}\gamma_5\right].$$
Similarly, a disconnected contribution to a meson correlation function can be written as $$C(t,t)={\rm Tr}\left[Q^{1}_{uv}\right].$$

\section{Preliminary Results}
The preliminary results shown in this section were obtained on $2+1$ dynamical, anisotropic lattices (anisotropy $a_s/a_t=3.5$) generated by the Hadron Spectrum Collaboration, \cite{Lin:2008pr}. Two values of the quark mass were used corresponding to pion masses of $\sim390$ MeV ({\bf 390 lattice}) and $\sim240$ MeV ({\bf 240 lattice}) with a spatial lattice spacing of $\sim0.12$ fm. Several lattice sizes were generated with most of our preliminary results from the $16^3\times 128$, $20^3\times 128$, $24^3\times 128$ lattices and the number of configurations ranging from $100\sim300$. 
\subsection{$I=2$}
\begin{figure}[h]
\centering
\includegraphics[width=110mm]{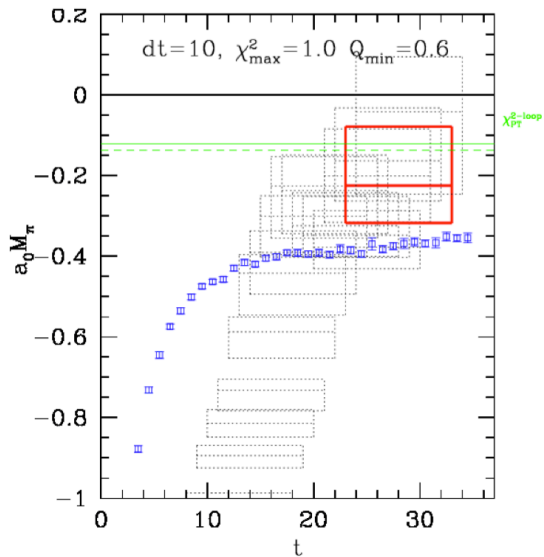}
\caption{A sliding window plot of $a_0M_\pi$ for the 240 data plotted together with the negative of the effective mass (to show the quality of the data in the fit region). The dashed boxes are fits where the $\chi^2_{dof}$ and $Q$ did not satisfy the condition, $\chi^2_{dof}<1$ and $Q>0.6$. The two-loop $\chi_{PT}$ value is shown with a solid horizontal line and the dashed horizontal line is an estimate of the error due to the lack of the measurement of $f_\pi$ on these lattices.} \label{fig:iso2effmass}
\end{figure}

The $I=2$ sector is an exotic sector without any disconnected diagrams or box diagrams. Its simulation, therefore, has the longest history as all of the diagrams can be evaluated using the traditional point-to-all quark propagators. The main advantage of using the stochastic LapH method in this case is to increase statistics on a given lattice and to include finite momentum operators in the variational basis. An example of the effective mass of this state is shown in Figure~\ref{fig:iso2effmass}. 

\subsection{$I=1$}
\begin{figure}[h]
\centering
\includegraphics[width=80mm]{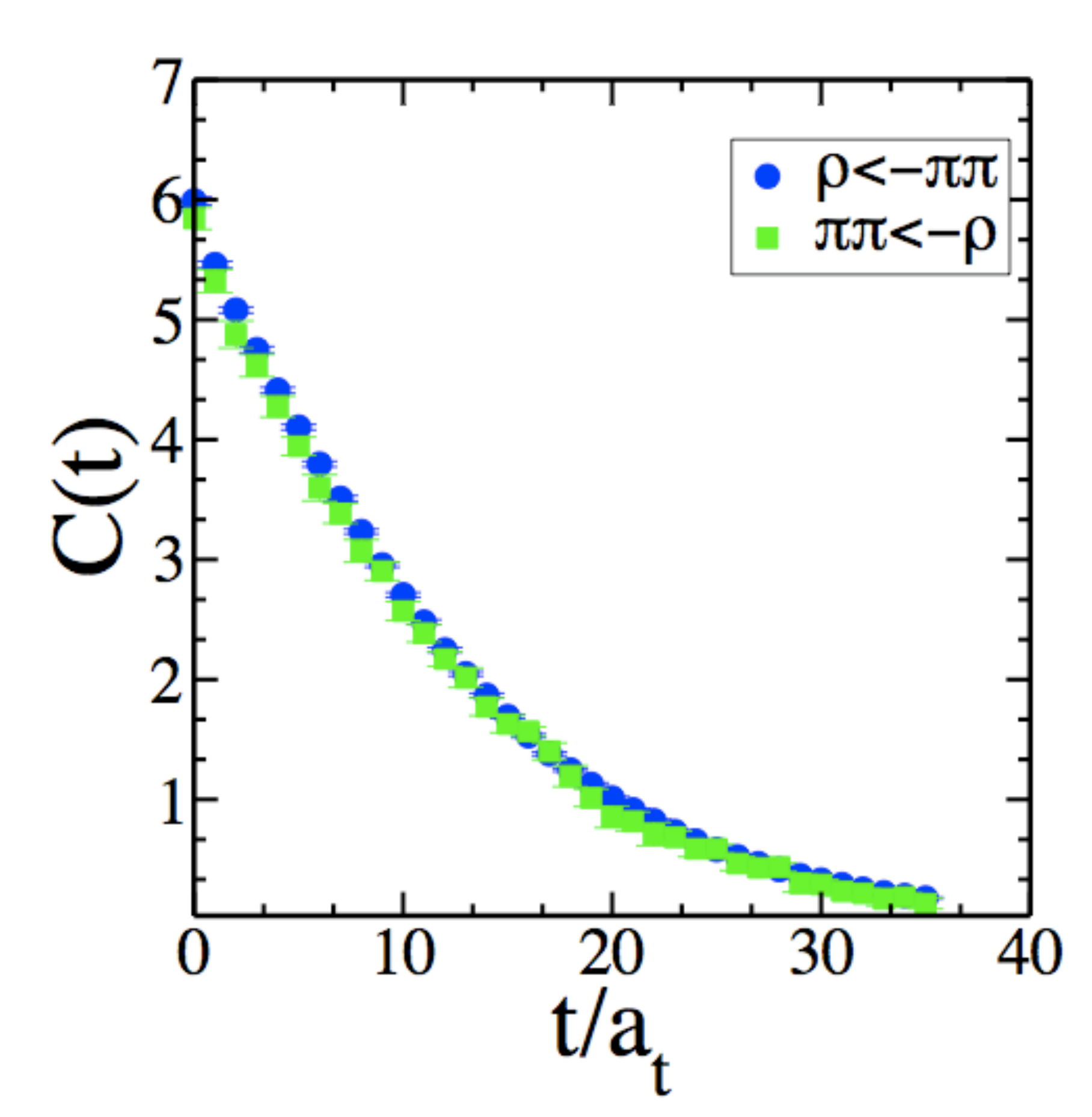}
\caption{The off-diagonal correlator of the $\rho\rightarrow\pi\pi$ mixing term.} \label{fig:rhopipi}
\end{figure}

In the $I=1$ sector, one must consider the mixing between the (single particle) rho and the two-pion state with sufficiently light quarks and large spatial volumes. The two pion diagram contains the $t$-$to$-$t$ term as one of the quarks are exchanged between the two final state pions. The correlation matrix with a two-pion operator and a single rho operator must be formed to determine the mixing strength and the mass of the ground state. the An example for the signal of this state is shown in Figure~\ref{fig:rhopipi}.

\subsection{$I=0$}
\begin{figure}[h]
\includegraphics[width=100mm]{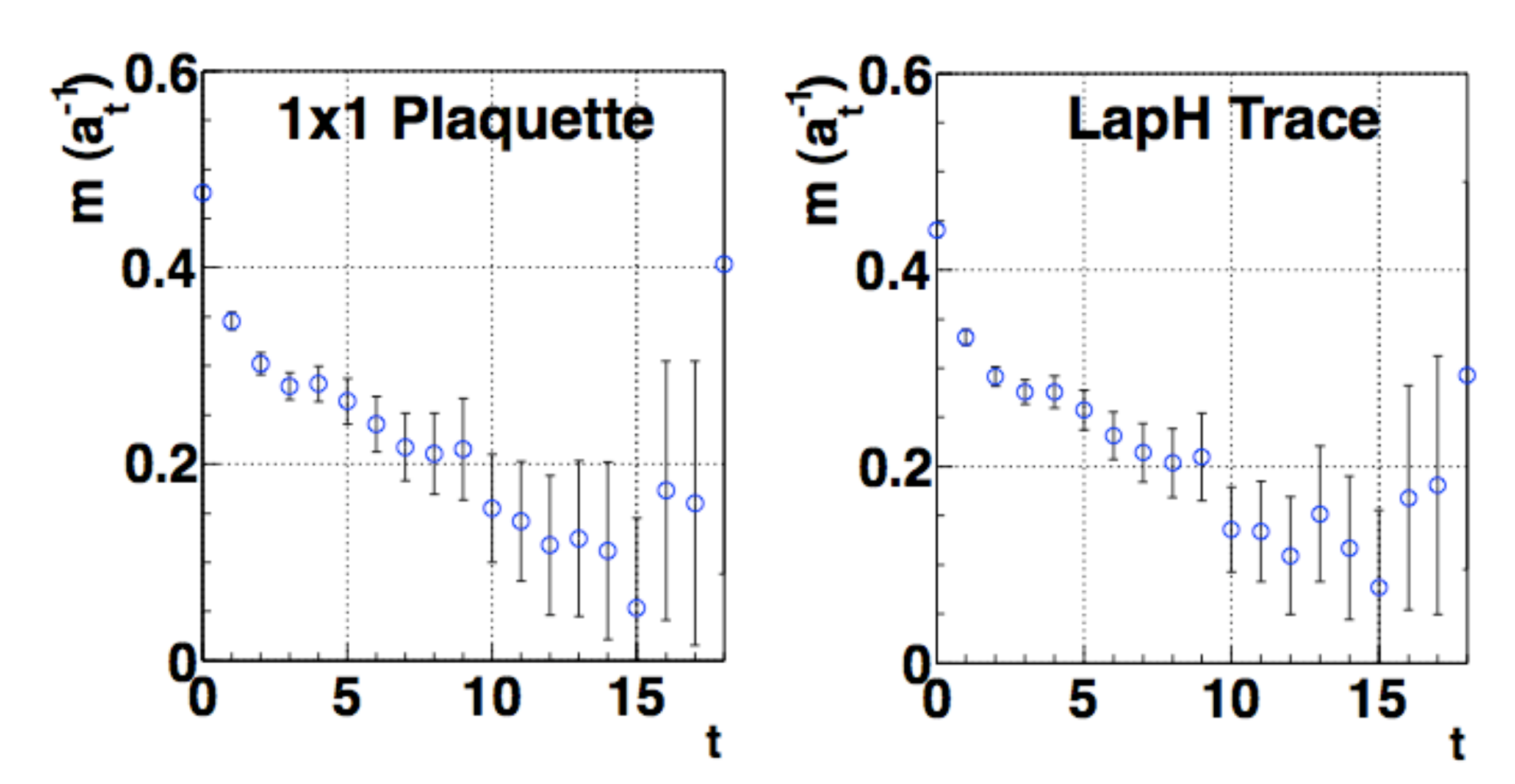}
\caption{Comparison of the standard plaquette operator and the LapH operator for the scalar glueball.} \label{fig:glueball}
\end{figure}

\begin{figure}[h]
\centering
\includegraphics[width=170mm]{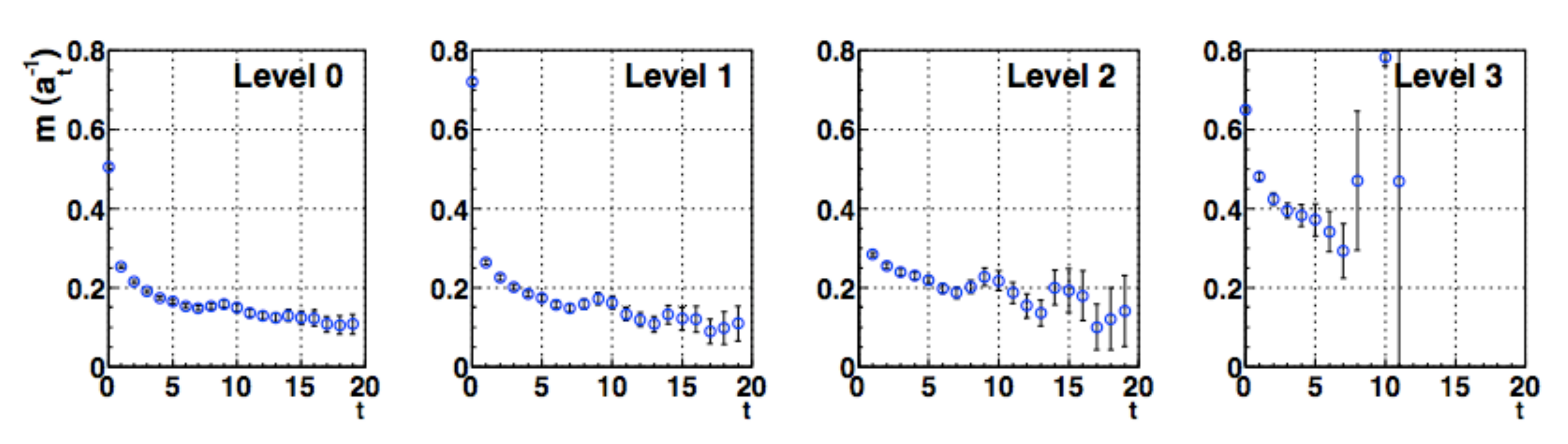}
\caption{The optimized, diagonal correlators in the $I=0$ sector using a glueball, meson and multi-particle operators as the basis of operators in the variational basis.} \label{fig:Ieq0}
\end{figure}

The $I=0$ pion-pion correlation function contain both disconnected diagrams and box diagrams (which have $t$-$to$-$t$ propagation on two of the ``end timeslices"). The disconnected diagrams have much smaller magnitude than the connected diagrams (in this case) and are usually noisier than the connected diagrams computed with the same number of configurations. The problem in this channel is even worse as the 
complete study of the isosinglet channel requires that we use a scalar glueball operator, quark-antiquark isoscalar meson-like operators, two-pion scattering operators and then study the mixing of the various components in this channel. The scalar glueball is one of the noisiest states (in the pure Yang-Mills theory) and the reliable extraction of the mixing with quark states has been a difficult problem to solve for some time. Since the stochastic LapH algorithm is a quark smearing algorithm, one may not immediately expect to gain anything in the glueball channel. However, the trace of the LapH operator has all of the symmetries needed for the $0^{++}$ channel and can in principal be used as a glueball operator. Figure~\ref{fig:glueball} shows the comparison of the effective masses between the usual single plaquette glueball operator and the LapH glueball operator. One sees that the LapH operator is just as good as the plaquette operator (or even better) for the scalar glueball. This implies that we can study all three types of $I=0$  operators with the stochastic LapH algorithm. The signal-to-noise is such that one can diagonalize this matrix to obtain the low-lying spectrum in the $0^{++}$ channel (Fig.~\ref{fig:Ieq0}).

\section{Summary}
A new quark smearing algorithm for lattice spectroscopy was presented with preliminary results in the isospin 0, 1 and 2 channels. A simple single exponential fit of the correlator of the state projected with two $p=0$ pion operators at $M_\pi=240$ MeV gave results which were consistent with continuum chiral perturbation theory, \cite{Colangelo:2001df}. Scattering states with finite momenta have also been obtained in the $I=2$ channel at pion masses close to $390$ MeV and work is under way to compute the phase shift with lighter pions and larger volumes. This will enable us to directly compare phase shifts as well as threshold parameters. We have calculated mixing diagrams between the single-particle $\rho$ operator and the two-pion scattering operators which show levels of noise that gives us confidence that decays can be studied with the stochastic LapH algorithm despite the presence of a $t$-$to$-$t$ propagator. The most surprising result, however, may be in the isoscalar sector where the trace of the LapH operator have allowed us to compute the mixing of a glueball operator, a scalar meson operator and two-pion operators with relatively low statistics. The preliminary results presented here on relatively small lattices strongly suggests that similar quality can be obtained at larger lattices and even lighter pion masses. We are currently generating the propagators needed to determine the low-lying spectrum of baryons and mesons with the correct treatment of the states above thresholds. 

\begin{acknowledgments}
This work was supported by the U.S. National Science Foundation under awards PHY-0510020, PHY-0653315, PHY-0704171, PHY-0969863, and PHY-0970137, and through TeraGrid resources provided by the Pittsburgh Supercomputer Center, the Texas Advanced Computing Center, and the National Institute for Computational Sciences under grant numbers TG-PHY100027 and TG- MCA075017. MJP is supported by Science Foundation Ireland under research grant 07/RFP/PHYF168. The USQCD QDP++/Chroma library (\cite{Edwards:2004sx}) was used in developing the software for the calculations reported here.
\end{acknowledgments}

\bigskip 


\end{document}